\documentclass[twocolumn,showpacs,superscriptaddress,amsmath,amssymb,prl]{revtex4}

\usepackage{graphicx}%
\usepackage{dcolumn}%

\begin{document}

\title{\textquotedblleft Waterfalls{\textquotedblright} phenomenon in superconducting cuprates}

\author{D. S. Inosov}
\affiliation{IFW Dresden, P.O. Box 270116, D-01171 Dresden, Germany}

\author{A. A. Kordyuk}
\affiliation{IFW Dresden, P.O. Box 270116, D-01171 Dresden, Germany}
\affiliation{Institute of Metal Physics of National Academy of Sciences of Ukraine, 03142 Kyiv, Ukraine}

\author{S. V. Borisenko}
\author{V. B. Zabolotnyy}
\affiliation{IFW Dresden, P.O. Box 270116, D-01171 Dresden, Germany}

\author{J.~Fink}
\affiliation{IFW Dresden, P.O. Box 270116, D-01171 Dresden, Germany}
\affiliation{BESSY GmbH, Albert-Einstein-Strasse 15, 12489 Berlin, Germany}

\author{M. Knupfer}
\author{B. B\"uchner}
\affiliation{IFW Dresden, P.O. Box 270116, D-01171 Dresden, Germany}

\author{R. Follath}
\affiliation{BESSY GmbH, Albert-Einstein-Strasse 15, 12489 Berlin, Germany}

\author{V. Hinkov}
\author{B. Keimer}
\affiliation{Max-Planck Institut f\"ur Festk\"orperforschung, 70569 Stuttgart, Germany}

\author{H. Berger}
\affiliation{Institut de Physique de la Mati\'ere Complexe, EPFL, 1015 Lausanne, Switzerland}

\date{March 8, 2007}%

\begin{abstract}
We show that the \textquotedblleft waterfalls", as reported in recent ARPES studies on HTSC, can neither be described as a part of a self-consistent quasiparticle spectrum nor represent a new physical phenomena, namely the \textquotedblleft new energy scale". They stem from the critical suppression of the photoemission intensity along the Brillouin zone (BZ) diagonals. Our arguments, however, do not question the existence of the high-energy scale itself ($\sim$ 0.25 eV), which is a simple consequence of the renormalization maximum and has been explained earlier in terms of coupling to a continuum of bosonic excitations. Moreover, when the matrix-elements are taken into account, it becomes clear that the photoemission spectrum consists of two components: one represents the spectrum of one-particle excitations and the other, having a grid-like structure along the bonding directions in the BZ, is of yet unknown origin.
\end{abstract}


\preprint{\textit{xxx}}

\maketitle

\begin{figure*}
\includegraphics[width=15cm]{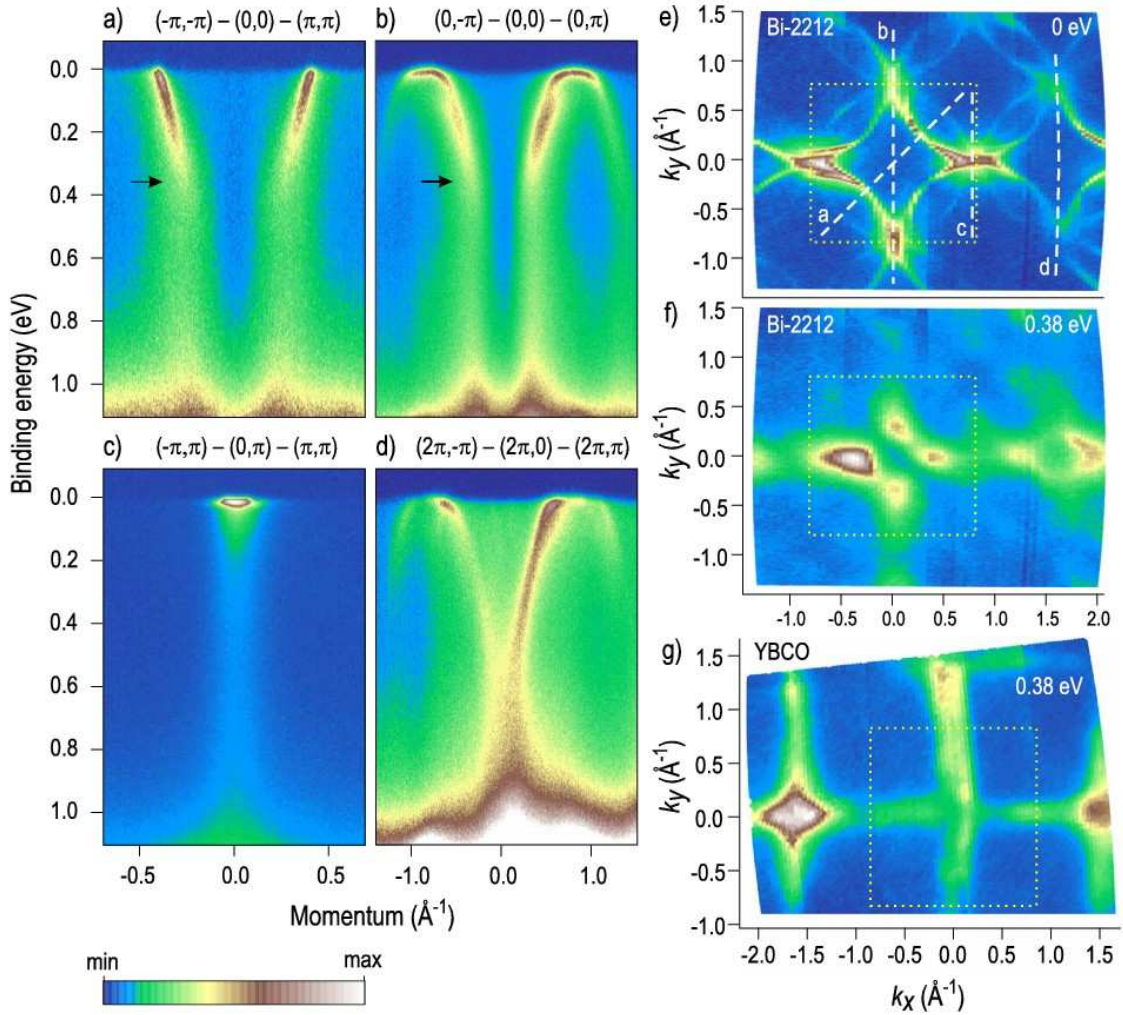}
\caption{\label{Fig1} Typical snapshots of the one-particle excitation spectrum of (Bi,Pb)$_2$Sr$_2$CaCu$_2$O$_8$ (a-f) and YBa$_2$Cu$_3$O$_{6.8}$ (g) as seen by angle resolved photoemission. The spectra shown in panels a-d are measured along the cuts marked on the Fermi surface map (e). f, g The distribution of the spectral weight at 0.38 eV below the Fermi level for BSCCO and YBCO samples, respectively. The 1st Brillouin zone is confined by the dotted squares on the maps.}
\end{figure*}

\begin{figure*}
\includegraphics[width=12cm]{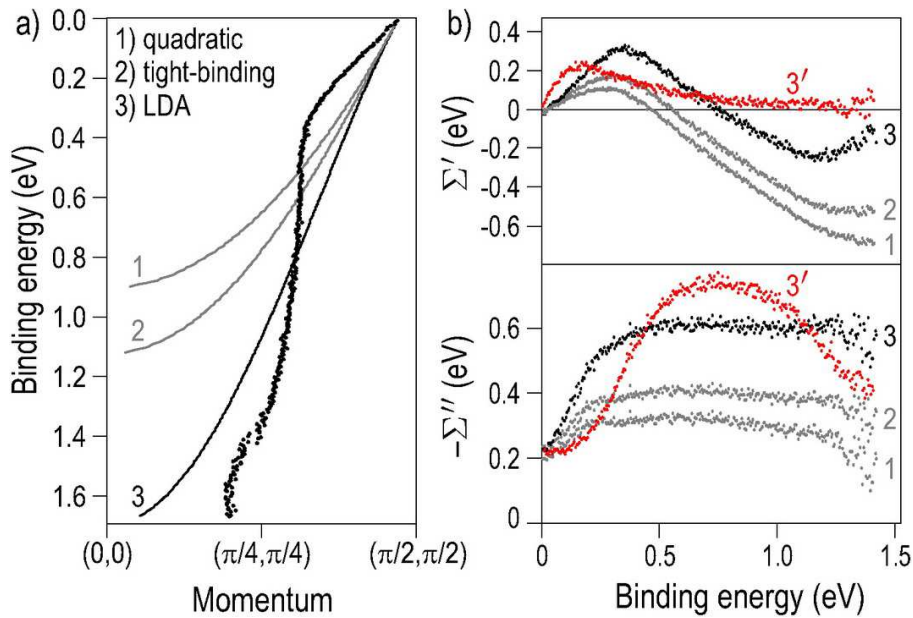}
\caption{\label{Fig1} The self-energy parts extracted from the waterfall spectrum (such as shown in Fig.1a) are inconsistent with Kramers-Kronig (KK) transform. (a) Experimental dispersion (points) with different bare band models: parabolic (1) \cite{17}, tight binding (2) \cite{19}, and LDA calculated (3) \cite{6}. (b) Real (top) and imaginary (bottom) parts of the self-energy extracted from experimental dispersion and from width of photoemission spectrum, respectively. Different curves (1--3) correspond to different bare band models. The red curves ($3'$) in both panels represent the KK-consistent counterparts for the LDA bare band model (see Ref.\;\onlinecite{17} for details of KK procedure).}
\end{figure*}

\begin{figure*}
\includegraphics[width=16cm]{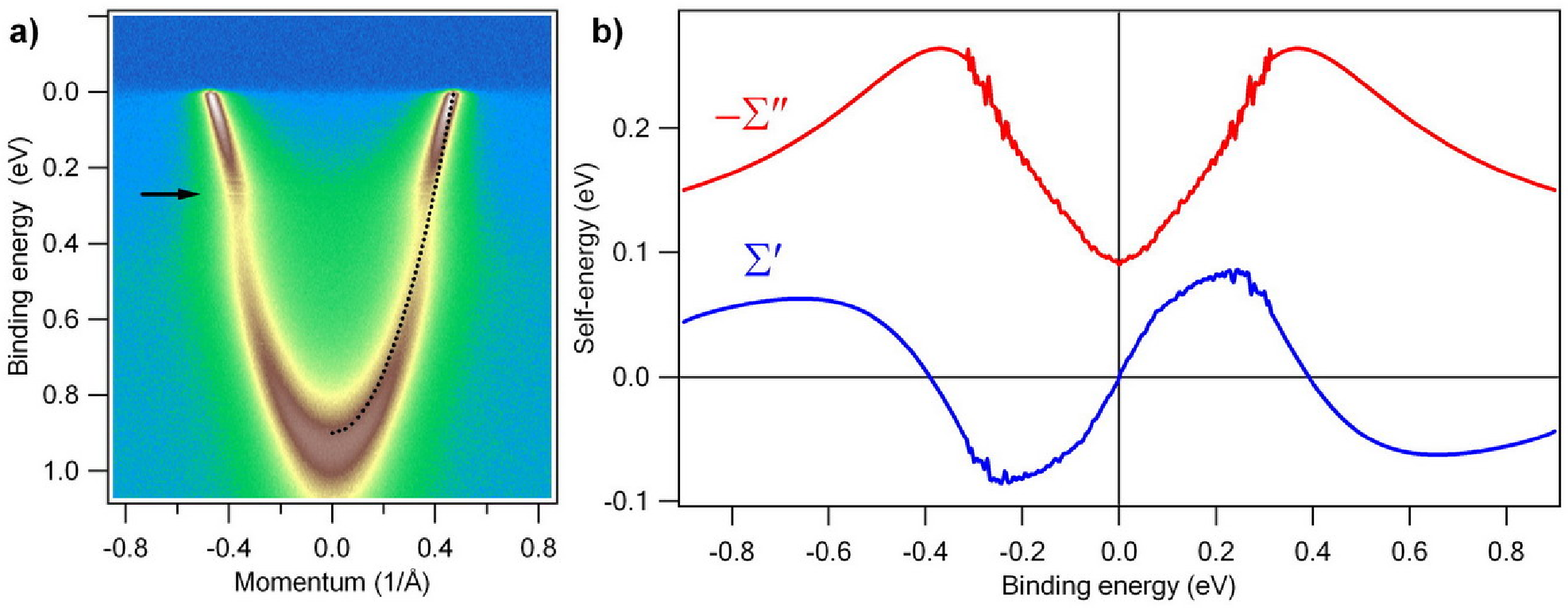}
\caption{\label{Fig3} The quasiparticle spectrum along the nodal direction of Bi-2212 (a) simulated from the complex self-energy (b), and bare electron dispersion (dotted line), both determined from the self-consistent analysis of the photoemission spectrum of Bi-2212 within $0-0.3$~eV binding energy range (after Ref.\;\onlinecite{17}).}
\end{figure*}

Since the mechanism of superconductivity in cuprates remains an unresolved issue, the observation of the relevant energy scale in their electronic excitation spectrum is naturally of great importance. Recently, a renewed interest to this problem has been rekindled by the finding of a \textquotedblleft high energy anomaly". This anomaly, termed \textquotedblleft waterfalls", is observed in the photoemission spectra of cuprates by several photoemission groups \cite{1,2,3,4,5,6} and already adopted by a number of different theories \cite{7,8,9,10,11,12,13}. Here we show that the high energy scales, obtained if one treats the waterfalls as a self-energy effect, may be wrong---caused by strong suppression of the photoemission intensity in the center of the Brillouin zone (BZ). Taking this suppression into account reveals a new additional component in the photoemission spectrum. This component exhibits a grid-like structure along the bonding directions of the BZ which may be a signature of a one-dimensional ordering in cuprates.

The waterfalls are observed as extended vertical parts of the quasiparticle dispersion around the BZ center \cite{1,2,3,4,5,6} (see Fig.1a,b) and explained in terms of new ideas such as a disintegration of the quasiparticles into a spinon and holon branch \cite{1}, coherence-incoherence crossover \cite{4,5}, disorder-localized band-tailing \cite{13}, by familiar $t$$-$$J$ model with \cite{7} or without \cite{9} string excitations, as well as within the self-energy approach by strong local spin correlations \cite{2}, itinerant spin fluctuations \cite{3,7}, or quantum-criticality \cite{12}. The reported distribution of the waterfalls' tails in momentum forms a diamond like shape \cite{1,2} around the BZ center. However, similar extended tails are also observed close to the BZ boundary at the ($\pi$,0) point \cite{2,5} (see Fig.\;1c), for example. 

First, we show that the waterfalls do not necessarily represent a new physical phenomena---the \textquotedblleft new energy scale". This becomes clear from comparison of the raw data presented in Fig.\;1 b and d. These panels represent two photoemission spectra measured along two equivalent cuts of the reciprocal space of a single-crystalline BSCCO, but while panel b shows the distinct energy scale (black arrow) and long waterfalls, very similar to Fig.\;1a, panel d exhibits virtually neither such a scale nor the waterfalls. Since both the electronic band structure and many body effects stay invariant to any translation by a reciprocal lattice vector, the difference between these two images comes from the photoemission matrix elements which, as a rule, strongly depend on momentum \cite{14,15,16}. In this case the photoemission is completely suppressed in the center of the first BZ or, more precisely, along the zone diagonals (nodal directions, see Fig.\;1e). One can see that both the anomalously sharp kink and the long vertical waterfalls in Fig.\;1b are simply caused by such a suppression.

Second, we address the question whether the waterfalls observed along the nodal direction (Fig.\;1a) can be described by a single-particle Green's function (as suggested by Refs.\;\onlinecite{2, 3, 6}) and are, therefore, self-consistent. Evidently, they are not. The \textquotedblleft self-consistency" means that the spectrum can be described by the Green's function $G(\mathbf{k},\omega) = 1/(\omega - \varepsilon_{\mathbf{k}} - \Sigma)$ where the quasiparticle self-energy  $\Sigma(\mathbf{k},\omega)$ is a casual analytic function, the real and imaginary parts of which are related by the Kramers-Kronig (KK) transform (see Ref.\;\onlinecite{17} for further details). The lack of self-consistency in the wide energy range, such as covered by Fig.\;1a or b, can be easily understood: while the imaginary part, $\Sigma''(\omega)$, which is proportional to the width of the quasiparticle spectrum in momentum (so-called \textquotedblleft MDC width" \cite{18}), stays constant along the waterfall length ($0.4-0.9$ eV), the real part, $\Sigma'(\omega)$, which is the difference in energy between the experimental and the bare dispersions, becomes inevitably negative when the long vertical waterfall crosses the bare dispersion. This is clearly seen in Fig.\;2, which also shows that not only the long sections of constant $\Sigma''(\omega)$ but also the energy scales derived from width and dispersion of the waterfall anomaly are not KK-consistent. For example, the maximum of  $\Sigma'(\omega)$ derived from the experimental dispersion (see curves 1-3 in top of panel b) stays at much higher energy, $\sim$ 0.36 eV, than the position of the maximum of $\Sigma'(\omega)$ = KK $\Sigma''(\omega)$ (curve $3'$), $\sim$ 0.17~eV.

Third, we note that the given arguments do not doubt the existence of the high-energy scale itself, which is a simple consequence of the renormalization maximum and has been already explained in terms of coupling to a continuum of bosonic excitations \cite{17,20,21,22,23}. Fig.\;3 presents a self-consistent quasiparticle spectrum which is simulated on the basis of the complex self-energy function and bare electron dispersion, both derived from the photoemission data measured up to binding energy of 0.3 eV and published in Ref.\;\onlinecite{17}. The high energy scale at about 0.25 eV (shown by the black arrow in Fig.\;3a) appears as a natural consequence of the maximum in $\Sigma''(\omega)$. Thus, it seems to be a matter of a pure coincidence that the high-energy scale at about 0.4 eV, caused by the aforementioned suppression, almost coincides with the real one, caused by the renormalization maximum.

Finally, we note that the observed suppression does not explain the whole story---i.e. why the waterfalls are so long. To better perceive the problem, the reader may compare the waterfalled photoemission spectrum (see Fig.\;1a or Refs.\;\onlinecite{1,2,3,4,5,6}) with the self-consistent one (see Fig.\;3a or Ref.\;\onlinecite{10}). Moreover, the problem of the long tails is not peculiar merely to the region around the BZ center. The long tail of constant intensity at the $(\pi,0)$ point, shown in Fig.\;1c, also cannot be described by self-consistent Green's function, since for $|\omega| \gg [\varepsilon_{\mathbf{k}}, \Sigma'(\omega), \Sigma''(\omega)]$ the one-particle spectral function $A \sim Im G \sim \Sigma''(\omega)/\omega^2$ and can hardly stay constant. On the other hand, these tails can find a natural explanation as caused by an additional spectral weight of yet unknown origin. Then the waterfalls seen in panels a and b are formed because the suppression affects both the quasiparticle (the spectral weight which can be described by the quasiparticle approach) and the additional (extrinsic or incoherent) components, as can be seen in Fig.\;1f. Here we can mention that an additional higher energy component is supported by recent optical experiments \cite{24}.

It is interesting that the distribution of this additional component in momentum space, when the photoemission matrix elements are taken into account, is localized along the bonding directions in the reciprocal space, as one can see in Fig.\;1f and, more clearly, in Fig.\;1g for an untwinned YBCO, for which the overall picture is qualitatively the same. Evidently, this component, either incoherent or extrinsic, represents a new phenomenon which should be understood. One possible explanations can be related with the disorder-localized in-gap states \cite{13}. The inelastic scattering of photoelectrons \cite{25} can be another option. On the other hand, the grid-like momentum distribution of this additional spectral weight may hint at the presence of a one-dimensional structure \cite{26}. Then, the photoemission spectra consists of two signals, one from the well studied two-dimensional metallic phase and another from an underdoped one-dimensional phase. Such a scenario would be consistent with the \textquotedblleft checkerboard" structure observed by scanning tunneling spectroscopy in lightly hole-doped cuprates \cite{27}.

We acknowledge the discussions with A. S. Alexandrov, D. J. Scalapino, T. Timusk, Binping Xie, and Donglai Feng. The project is part of the Forschergruppe FOR538. The work in Lausanne was supported by the Swiss National Science Foundation and by the MaNEP.

\end{document}